\documentclass[twocolumn,showpacs,amssymb,prl]{revtex4}

\usepackage{graphicx}% Include figure files
\usepackage{dcolumn}% Align table columns on decimal point
\usepackage{bm}% bold math
\def\beq{\begin{equation}}
\def\enq{\end{equation}}

\def\eps{\epsilon}

\begin{document}
\input{epsf}

\title{High Energy Neutrinos from Gamma-Ray Bursts with Precursor
Supernovae}

\author{Soebur Razzaque,$^1$ Peter M\'esz\'aros$^1$ and Eli
Waxman$^2$}

\affiliation{$^1$Department of Astronomy \& Astrophysics, Department
 of Physics, Pennsylvania State University, University Park,
 Pennsylvania 16802, USA \\ $^2$Department of Condensed Matter
 Physics, Weizmann Institute of Science, Rehovot 76100, Israel}

\begin{abstract}
The high energy neutrino signature from proton-proton and photo-meson
interactions in a supernova remnant shell ejected prior to a gamma-ray
burst provides a test for the precursor supernova, or supranova, model
of gamma-ray bursts.  Protons in the supernova remnant shell, and
photons entrapped from a supernova explosion or a pulsar wind from a
fast-rotating neutron star remnant provide ample targets for protons
escaping the internal shocks of the gamma-ray burst to interact and
produce high energy neutrinos.  We calculate the expected neutrino
fluxes, which can be detected by current and future experiments.

\end{abstract}

\date{\today}
\maketitle

Gamma-ray bursts (GRBs) are thought to be possible sources of high
energy neutrinos. In the currently favored models the $\gamma$-ray
emission is attributed to radiation from shock-accelerated electrons
in the relativistic fireball outflow or jet from a cataclysmic stellar
event. The latter may be connected to compact stellar mergers, or the
core collapse of a massive stellar progenitor (collapsar), which could
also involve a core-collapse supernova (SN) \cite{woo93}.  Together
with electrons, the shocks are expected to accelerate protons as well,
and high energy neutrinos are thought to be produced dominantly by
$p\gamma$ interactions of the protons with synchrotron or inverse
Compton scattered photons \cite{wb97,wb00}. Recent reports of
detection of X-ray lines from several GRB afterglows have been
interpreted \cite{piro00}, although not unambiguously, as providing
support for a version of the collapsar model in which a SN explosion
occurs weeks before the GRB (the ``supranova" model, \cite{vie98}).
In the supranova scenario the supernova remnant (SNR) shell provides
nucleon targets for $pp$ interactions with protons accelerated in the
MHD wind of a pre-GRB pulsar \cite{gg02}, leading to a 10 TeV neutrino
precursor to the GRB (other nucleon targets from a stellar companion
disruption leading to $\pi^0$ decay GeV $\gamma$-rays were discussed
by \cite{katz94}). The SNR also provides additional target photons for
$p\gamma$ interactions \cite{gg02} with internal shock-accelerated
protons, resulting in $\sim 10^{16}$ eV neutrinos.  In this Letter we
investigate some important unexplored aspects of the GRB-SNR
interaction, namely $pp$ and $p\gamma$ interactions involving GRB
protons in the shocked shell, which have significant consequences in
the assessment of the neutrino signatures from these objects.

{\it SNR target nucleons and photons.---} The typical mass ejected in
a SN is $M_{\rm snr} \sim m_{\rm snr} M_{\odot} = 2 \times 10^{33}
m_{\rm snr}$ g. For a nominal sub-relativistic shell speed $v=10^9$
cm/s, the typical distance reached is $R_{\rm snr} \sim 10^{14} v_9
t_d$ cm in $t_d$ days. We assume that the gas is roughly isotropically
ejected in a shell of width $\delta= \Delta R /R = 0.1 \delta_{-1}$
and average column density $\Sigma \approx 1.6\times 10^4 m_{\rm snr}
v_9^{-2} t_d^{-2}$ g-cm$^{-2}$.

The deposition of $10^{51}E _{51}$ ergs in a SN progenitor stellar
envelope of dimension $R_\ast=10^{12}R_{12}$ cm heats it to a
temperature $T_{\rm o} \sim 1\,E_{51}^{1/4}R_{12}^{-3/4}
\delta_{-1}^{-1/4}$ keV. The mean photon energy in the SNR shell is $
\eps_{\gamma,{\rm sn}} \sim 20\, E_{51}^{1/4}\delta_{-1}^{1/3}
R_{12}^{1/4}v_9^{-1} t_d^{-1} ~{\rm eV}$.  The photon column density
in the SNR is $\Sigma_{\gamma,{\rm sn}} \sim 3\times 10^{32}
E_{51}^{3/4}\delta_{-1}^{1/3}R_{12}^{-1/4}v_9^{-1} t_d^{-1}$
cm$^{-2}$.  The $p\gamma$ optical depth and threshold proton energy at
the $\Delta$-resonance are
\begin{eqnarray}
\tau_{p\gamma,{\rm sn}} &\sim & 3\times 10^4
E_{51}^{3/4}\delta_{-1}^{1/3}R_{12}^{-1/4}v_9^{-1} t_d^{-1} 
\nonumber \\ 
E_{p,{\rm th, sn}} &\sim & 10^7
E_{51}^{-1/4}\delta_{-1}^{-1/3}R_{12}^{-1/4}v_9 t_d ~\hbox{GeV.}
\label{eq:taupgsn}
\end{eqnarray}

The collapse of the Fe core can lead initially to a fast-rotating
pulsar of rotational energy $E_{\rm rot}\sim 10^{53}$ erg and
spin-down time $\sim 3\times 10^6$ s \cite{ostgun69,gg02}, emitting an
MHD wind of luminosity $L_{\rm m}\sim 10^{46} L_{\rm m46}$ erg/s.
Besides ejecting the SNR envelope, the SN explosion may leave behind
parts of the He core, which take longer in falling back to make a
black hole leading to the GRB (months in the supranova scenario).  If
the fall-back is anisotropic and a channel forms in the He core
(e.g. along the rotation axis), the relativistic MHD wind may flow out
and impact the SNR shell further out. The MHD wind is highly
relativistic, and is decelerated to a subrelativistic velocity in a
reverse shock, driving a forward shock into the shell \cite{rg74}.
%(This differs from \cite{gg02}, who consider quasi-monoenergetic 
%protons carried by the wind, which impact the shell). 
The shock velocity $v_s$ in the observer frame satisfies, from 
pressure balance in the shock frame, $v_s-v\simeq [L_{\rm m}
(10^{-1}R/c)/M_{\rm snr}]^{1/2} \simeq 10^{-1}[E(t)/M_{\rm
snr}]^{1/2}\sim 10^{-2}c$ for typical $E(t)=tL_{\rm m}$ and $M_{\rm
snr}$. Thus the shock will propagate through a significant fraction 
of the shell width, and we approximate the situation by taking the 
shock to cross the entire shell, producing photons distributed 
throughout the shell.  Given the high Thomson optical depth in the 
SNR shell, these photons thermalize to a maximum energy
$\eps_{\gamma,{\rm m}} \sim 0.1 \, L_{\rm m46}^{1/4}
\delta_{-1}^{1/4}v_9^{-3/4} t_d^{-1/2}~ \hbox{keV}$,
assuming that all the MHD wind energy goes into photons.  The
$p\gamma$ optical depth and threshold proton energy at
$\Delta$ production, are
\begin{eqnarray}
\tau_{p\gamma,{\rm m}} &\sim & 3\times 10^4 L_{\rm m46}^{3/4}
\delta_{-1}^{1/4} v_9^{-1/2}t_d^{-1/2} \nonumber \\
E_{p,{\rm th,m}} &\sim & 2\times 10^6 L_{\rm m46}^{1/4}
\delta_{-1}^{-1/4} v_9^{-3/4}t_d^{-1/2}~\hbox{GeV}.
\label{eq:taupgm}
\end{eqnarray}

Additional target photons may arise from parts of the disrupted He
core, no longer in hydrostatic equilibrium and moving outwards inside
the shell. However, even if they have super-Eddington luminosities
$L_{\ast, {\rm He}} \sim 10^{44}$ erg/s as inferred in some SN, these
photons have an optical depth $\tau_{p\gamma}\sim 5 L_{44} v_9^{-2}
t_d^{-1}$ for protons with $E_p \gtrsim 3\times 10^8 t_d$ GeV,
negligible compared to Eqs. (\ref{eq:taupgsn}) and (\ref{eq:taupgm}).

The non-relativistic MHD forward shock moving into the SNR is likely
to be collisionless (as also generally assumed for non-relativistic
GRB internal shocks \cite{m02}). The transition is dominated by the
process giving the narrower shock width, i.e. the shortest timescales
for changing particle momenta. A collisionless shock thickness is
$\sim c/(\beta \omega_p)$, where $\beta$ is shock velocity, $\omega_p$
is plasma ion frequency. For our parameters $\sim 1~M_\odot$ shell at
$\sim 10^{14}$ cm radius, $\omega_p=3\times 10^{10}$/sec and
$\beta=0.01$, this shock width is 100 cm.  The thickness of a
radiation shock is $l/\beta$, where $l$ is the photon mean free path,
giving a thickness $\sim 10^{11}$ cm. The thickness for $pp$
collisions is even larger due to a smaller cross section than
Thomson. This strongly suggests that the shock is collisionless.

The average magnetic field inside the SNR shell is
\beq
B\sim 10^4 \xi_B^{1/2} m_{\rm snr}^{1/2} \delta_{-1}^{-1/2} v_9^{-1/2}
t_d^{-3/2} \;{\rm G},
\label{eq:Bshell}
\enq
where $\xi_B$ is the equipartition value in respect to the proton
kinetic energy. Equating the acceleration time $t_a\sim 2\pi
Amc\gamma/(\beta^2 eB)$ [where $\beta=(v_s-v)/c \sim
10^{-2}\beta_{-2}$ is the relative shock speed in the shell, and $A
\sim 10A_{1}$] to the synchrotron loss time $t_{\rm sy}=3m^3 c^5/(2
e^4B^2\gamma)$, the maximum accelerated particle Lorentz factor is
$\gamma_{\rm mx}=2\times 10^{10} A_1^{-1/2}(m/m_p)B^{-1/2}$. For the
field of Eq. (\ref{eq:Bshell}), the maximum proton Lorentz factor is
$\gamma_{p, {\rm mx}} \sim 2\times 10^7 A_1^{-1/2}\beta_{-2}
\xi_B^{-1/4}m_{\rm snr}^{-1/4} \delta_{-1}^{1/4}v_9^{1/4}t_d^{3/4}$.
The maximum electron Lorentz factor is $m_e/m_p$ lower and the
synchrotron photon peak energy is $\omega_{\rm sy} \approx \sim 0.2
A_1^{-1}\beta_{-2}^2$ keV, but due to the high Thomson depth these
photons are thermalized to the blackbody value $\eps_{\gamma,{\rm m}}
\sim 0.1$ keV.

{\it GRB protons and interactions in the SNR.---} The GRB
isotropic-equivalent $\gamma$-ray luminosity is $L_{\gamma}^{\rm iso}
\approx 10^{52} L_{52}$ ergs/s.  The corresponding GRB total proton
luminosity is $dN_p/dt = L_\gamma/ \Gamma m_p c^2 = 2\times 10^{52}
L_{52} \Gamma_{300}^{-1}$ s$^{-1}$. Assuming that a fraction $\xi_p
\lesssim 1$ of these is accelerated in the internal shocks results in
a proton distribution $d^2N_p/dE_p dt \simeq 6\times 10^{54}
L_{52}\xi_p E_{p, {\rm GeV}}^{-2}~ {\rm GeV}^{-1} {\rm s}^{-1}$.  High
energy protons interact with synchrotron and inverse Compton scattered
photons in the GRB fireball shock dominantly through $p\gamma
\rightarrow \Delta \rightarrow n\pi^{+}/p\pi^0$ \cite{wb97}, resulting
in $\gtrsim 100$ TeV neutrinos.  The cross-section for $\Delta$
production at threshold $E_pE_{\gamma} \sim 0.2\Gamma^2$ GeV$^2$ in
the observer frame is $\sim 0.1$ mb.  The optical depth for $E_p >
10^{7}$ GeV is $\tau_{p\gamma} \gtrsim 1$ in the high $\sim$ MeV
photon density in the internal shocks.  Since protons lose $\sim 20\%$
of their energy per interaction, each proton will undergo a couple of
interactions, thus roughly half of the protons with $E_p>10^{16}$ eV
will be converted to neutrons and escape.  At lower energies, protons
may be prevented by fireball magnetic field from escaping to the SNR.
However, since internal shock radii range over $3\times
10^{12}-3\times 10^{14}$ cm, this may allow a significant fraction
of $\lesssim 10^{16}$ eV protons to interact with the SNR shell.  This
implies that a fraction $\eta_p(E_p) \lesssim 1$ of protons escape the
fireball shock region to propagate outwards. The isotropic-equivalent
observer-frame proton luminosity impacting the shell is
\begin{eqnarray} 
\frac{d^2N_p}{dE_p dt} = 6\times 10^{54} E_{p,{\rm
GeV}}^{-2} L_{52} \xi_p \eta_{p}(E_p)~ {\rm GeV}^{-1} {\rm s}^{-1}~.
\label{eq:dNdEdt}
\end{eqnarray}

The large $p\gamma$ optical depths for both the SN-shock $\eps_{\rm
sn}$ and the pulsar MHD wind shock $\eps_{\rm m}$ photons
[Eqs. (\ref{eq:taupgsn}) and (\ref{eq:taupgm})] trapped inside the SNR
shell causes most of the incoming protons above the thresholds
[Eqs. (\ref{eq:taupgsn}) and (\ref{eq:taupgm})] to undergo photomeson
production.  Low energy protons below $\Delta$ production threshold
may interact with SNR protons. If $\xi_{\rm sh}$ (typically $\lesssim
10^{-1}$) is the fraction of shell protons which is accelerated to
relativistic energies by the collisionless MHD shock going through the
SNR shell, the column density of cold protons in the SNR shell is
$\Sigma_p \sim 10^{28} \zeta_{\rm sh} m_{\rm snr} v_9^{-2} t_d^{-2}$
cm$^{-2}$, where $\zeta_{\rm sh} =1-\xi_{\rm sh}\sim 1$.

The total cross-section for $pp$ interaction has been measured up to
very high energies (120 mb at $\sqrt{s} = 30$ TeV \cite{hagiwara02})
in accelerator experiments.  We take the mean $pp$ cross-section to be
$<\!\! \sigma_{pp} \!\!> \approx 100 \;\mbox{mb}$ in the TeV and above
energy range.  The corresponding mean optical depth is
\begin{equation}
<\!\! \tau_{pp} \!\!> = \Sigma <\!\! \sigma_{pp} \!\!> \approx 10^3
\zeta_{\rm sh} m_{\rm snr} v_9^{-2} t_{d}^{-2} ~.
\label{ppoptical}
\end{equation}
To calculate the neutrino flux from $pp$ interactions, one needs to
know the secondary charged particle multiplicities as a function of
$\sqrt{s}$.  Analytic calculations of this multiplicity, known as the
KNO scaling law \cite{kno}, are based on the rapidity ($y = {\rm
ln}[(E+p_z)/(E-p_z)]/2$, taking the beam direction along the $z$-axis)
distribution of the secondary charged particles.  Analytically the
total charged particle multiplicity increases as $\sim {\rm ln} (s)$.
Accelerator data show a slightly faster growth in $pp$ interactions
measured up to $\sqrt{s} = 1.8$ TeV \cite{ua5e735}.  An extrapolation
up to $\sqrt{s} \sim 400$ TeV, relevant for our calculation, for
$\pi^{\pm}$ and $K^{\pm}$ multiplicities from $pp$ interactions using
$\sim {\rm ln} (s)$ scaling is certainly very conservative.  There
exist in the literature other faster growing models such as $\sim
s^{0.3}$ \cite{engel98}.  The dominant neutrino production channels in
our calculation are $pp \rightarrow \pi^{\pm}/K^{\pm} \rightarrow \mu
\nu_{\mu} \rightarrow e \nu_e \nu_{\mu} {\bar \nu}_{\mu}$.  High
energy photons are also produced from $\pi^0 \rightarrow
\gamma\gamma$.

We used the {\small PYTHIA 6.2} event generator \cite{pythia} to
simulate $pp$ interactions \cite{prep}.  The angular deviation of a
secondary particle is related to its Lorentz invariant rapidity ($y$)
as $\theta \approx 1/\cosh(y)$.  We select $\pi^{\pm}$ and $K^{\pm}$
which are forward, namely, $y\ge 0$.  The average $\pi^{\pm}$ and
$K^{\pm}$ multiplicities measured at $\sqrt{s}=540$ GeV in the
pseudo-rapidity ($\eta \approx y$ for momentum $p\gg m$) region $0
\leq \eta \le 5$ are $<\!\!  n_{\pi} \!\!> = 11.15$ and $<\!\! n_{K}
\!\!> = 1.25$ \cite{ua5e735}.  These numbers agree with our {\small
PYTHIA} simulations within error bars.  For $y \ge 0$ we have $<\!\!
n_{\pi} \!\!> = 15.3$ and $<\!\! n_{K} \!\!> = 1.7$ at $\sqrt{s}=540$
GeV from simulations.  At the highest energy, $E_p =10^{20}$ eV
($\sqrt{s}=4.33 \times 10^5$ GeV), ${\rm ln}(\sqrt{s})$ extrapolation
gives $<\!\! n_{\pi} \!\!> = 15.3 \times {\rm ln}(4.33\times 10^5/540)
\approx 103$, about 75\% of our simulated value of 136.  Similarly
$<\!\!  n_{K} \!\!> = 1.7 \times {\rm ln}(4.33\times 10^5/540) \approx
11.4$, also about 75\% of our simulated value of 15.2 at $E_p=10^{20}$
eV.  For our calculation we used more conservative values: $<\!\!
n_{\pi} \!\!> = 103$ and $<\!\!  n_{K} \!\!> = 11.4$ at $E_p =10^{20}$
eV and ${\rm ln}(\sqrt{s})$ scaling at lower energies.

Secondary charged particles ($\pi^{\pm}$ and $K^{\pm}$) in the $pp$
interaction follow a $1/E$ energy distribution \cite{prep}.  The
energy of a particle of mass $m$ ranges from $ m \cosh (y) \gamma_{\rm
cm}$, where $\gamma_{\rm cm}$ is the Lorentz boost of the center of
mass in the lab frame and $\cosh (y)$ is the rapidity dependent boost
factor, up to the primary proton energy $E_p$.  For $y\ge 0$ we use
for secondary $\pi^{\pm}$ and $K^{\pm}$ the range
\beq
\gamma_{\rm cm} m_{\pi,\,K} \le E_{\pi,\,K} \lesssim E_p~.
\label{piKrange}
\enq
The pions and kaons decay into neutrinos with Lorentz-expanded decay
times $(\tau_{\pi,K}\gamma_{\pi,K})$ in the observer frame.

In the $\pi^{\pm}/K^{\pm} \rightarrow e \nu_e \nu_\mu {\overline
\nu}_\mu$ decay the $e$ and 3 $\nu$'s each share roughly 1/4 of the
pion/kaon energy.  The branching ratio for the kaon decay channel is
$\sim 64\%$.  The $\nu$ multiplicity for an incident proton of energy
$E_p \le 10^{11}$ GeV, using Eq. (\ref{piKrange}), is then
\begin{eqnarray}
M_{\nu}(E_p) &=& \frac{1}{4} {\cal N}_{\pi,K} \left(
\frac{E_{\nu}}{\rm GeV} \right)^{-1} \left[ \frac{1}{2}{\rm ln}\left(
\frac{10^{11}\, {\rm GeV}}{E_p} \right) \right]^{-1} \nonumber \\
&& \times \; \Theta \left( \frac{1}{4} \frac{m_{\pi,K}}{\rm GeV}
\gamma_{\rm cm} \leq \frac{E_{\nu}}{\rm GeV} \leq \frac{1}{4
}\frac{E_p}{\rm GeV} \right)
\label{numult}
\end{eqnarray}
for each type of neutrinos: $\nu_e$, ${\overline \nu}_{\mu}$ and
$\nu_{\mu}$.  Here $\Theta$ is a step function following from
Eq. (\ref{piKrange}).  The normalization factors ${\cal N}_{\pi}$ and
${\cal N}_{K}$ are found by integrating the $1/E$ distribution of
$\pi^{\pm}$ and $K^{\pm}$ in the energy range given in
Eq. (\ref{piKrange}) for $E^{\rm mx}_{\pi,K} \approx 10^{11}$ GeV and
equating to the respective total numbers, ${\cal N}_{\pi} = <\!\!
n_{\pi} \!\!>/{\rm ln} \left(E^{\rm mx}_{\pi}/m_{\pi} \gamma_{\rm cm}
\right) \approx 7 $, ${\cal N}_{K} = 0.64 <\!\! n_{K} \!\!>/{\rm ln}
\left(E^{\rm mx}_{K}/m_{K}\gamma_{\rm cm} \right) \approx 0.6$.

An additional $pp$ component may result from protons accelerated by
the collisionless shock in the shell colliding with shell thermal
nucleons.  For a typical SNR shell energy $10^{51}E_{51}$ ergs this
neutrino component has $\lesssim \zeta_{sh} 10^{51}E_{51}$ ergs, at
least an order of magnitude below the GRB shock proton component, but
it could become important for exceptionally energetic supernova
shells.

{\it Neutrino flux calculation.---} The GRB protons from internal
shocks undergo $pp$ interactions below $E_{p,{\rm th}}$ from
Eqs. (\ref{eq:taupgsn}) and (\ref{eq:taupgm}), and $p\gamma$
interactions above $E_{p,{\rm th}}$.  The neutrino flux ($\Phi_{\nu} =
{d^2N}/{dE_\nu dt}$) at Earth from a single GRB-SNR at distance $D$ is
\begin{eqnarray}
\Phi_{\nu}  = \frac{1}{4 \pi D^2} \left\{ \begin{array}{ll} 
\int f_{pp} M_{\nu} (E_p) \frac{d^2N}{dE_p dt} dE_p; & E_p
\lesssim E_{p,{\rm th}} \\
(f_{\pi}/4) \frac{d^2N}{dE_p dt}; & E_p > E_{p,{\rm th}}
\end{array} \right.
\label{single_nuflux}
\end{eqnarray}
for each $\nu$ type.  Here $f_{pp} = {\rm min}(1, <\!\! \tau_{pp}
\!\!>)$ from Eq. (\ref{ppoptical}), and $f_{\pi}$ is the fraction of
$E_p$ lost to $\pi$'s in $p\gamma$ interactions \cite{wb97}, which is
$\sim 1$ for an SNR with $\tau_{p\gamma}\gg 1$.  The synchrotron
cooling energy $E_{\nu}^{\rm s}$ is found by equating the decay times
($\tau_j\gamma_j$ with $j=\pi,K,\mu$) and synchrotron cooling times
for $\pi$'s, $K$'s and $\mu$'s. For the magnetic field in
Eq. (\ref{eq:Bshell}) we get the maximum energies as
\begin{eqnarray}
\gamma_j^{\rm s} \approx 2.3 \times 10^{4} m_j^{3/2} \tau^{-1/2}_{j}
\xi_{B}^{-1/2} m_{\rm snr}^{-1/2} \delta_{-1}^{1/2} v_9^{1/2}
t_d^{3/2}
\label{syncooltime}
\end{eqnarray}
where $m_j$'s are in GeV and $\tau_j$'s are in s.  Taking $m_{\rm
snr}\sim 10$ and typical age $t_d \sim 50$ d, the maximum energies
from Eq.  (\ref{syncooltime}) are $E_{\pi}^{\rm s} \sim 10^{10}$ GeV,
$E_{K}^{\rm s} \sim 10^{11}$ GeV and $E_{\mu}^{\rm s} \sim 10^{9}$
GeV, where we used $\xi_B \approx 10^{-2}\xi_{-2}$ as in typical GRB
fits.  The corresponding $\nu$ steepening break energies are $\gtrsim
10^9$ GeV.  The cooling times for inverse Compton scattering are
longer than respective lifetimes in the Klein-Nishina limit for $t_d
\gtrsim 10$ d.

The detection probability for a muon neutrino in an ice
detector is ${\cal P} \approx 1.7 \times 10^{-6} (E_{\nu}/{\rm
TeV})^{\beta}$, where $\beta=0.8$ for $E_{\nu}<1$ PeV and $\beta =
0.36$ for $E_{\nu}>1$ PeV. Multiplying this with the flux
of Eq. (\ref{single_nuflux}) and the burst duration, a numerical 
integration gives the expected event rates in a km$^2$ detector
from a single GRB.  The total number of $\nu_{\mu}$-induced upward
muon events for a single GRB of luminosity $10^{52}\,L_{52}$ ergs/s
and burst duration $\Delta t =10 \Delta_1$ s at $D \approx 1.6\times
10^{28} h_{65}^{-1} [(1+z)/2-(1+z)^{1/2}/2^{1/2}]$ cm $\sim 10^{28.2}
D_{28.2}$ cm for redshift $z \sim 1$ is
\begin{eqnarray}
N_\mu^{\rm TeV-PeV} & = & 1.1 D_{28.2}^{-2} \Delta_{1} \xi_p \eta_p(E)
L_{52} \nonumber \\ 
N_\mu^{\rm PeV-EeV} & = & 0.06 D_{28.2}^{-2}\Delta_{1} \xi_p \eta_p(E)
L_{52} .
\label{single_muevents}
\end{eqnarray}
Here we assumed $f_{pp} = 1$ and $f_{\pi} \sim 1$ for $m_{\rm snr}
\sim 10$ and $10 \lesssim t_d \lesssim 100$ days. These single-burst
numbers can be $\sim 10^2$ times larger, e.g. for rare bursts with
$D=10^{27.5}$ cm and $\Delta t=100$ s occurring at rates $\sim$ 3
yr$^{-1}$.

\begin{figure}
\centerline{\epsfxsize=3.4in \epsfbox{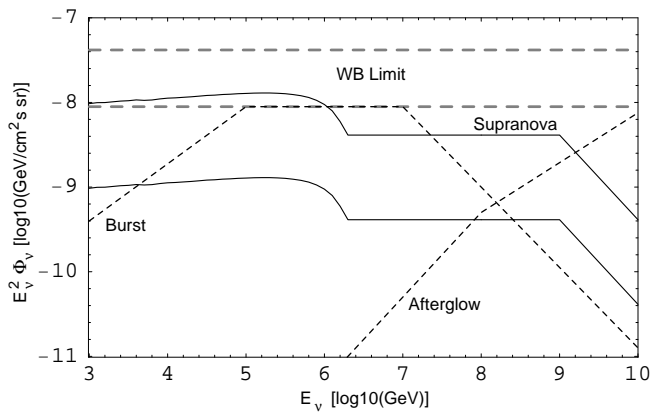}}
\caption{Diffuse neutrino flux ($E_{\nu}^2 \Phi_{\nu}$) from
post-supernova (supranova) models of GRBs (solid curves), assuming
that (top curve) all GRBs have an SNR shell, or (bottom) 10\% of all
GRBs have an SNR shell, and $10^{-1}$ of the fireball protons reach
the shells.  Long dashed lines correspond to the Waxman-Bahcall
cosmic-ray limit, short dashed curves are the diffuse $\nu$ flux from
GRB internal shocks and afterglows.}
\label{fig:difflux}
\end{figure}

The diffuse $\nu$-flux is obtained from Eq. (\ref{single_nuflux})
multiplied by $\Delta t$ and an observed rate $\sim 600/4\pi$
yr$^{-1}$ sr$^{-1}$.  Figure \ref{fig:difflux} shows the diffuse
$\nu_\mu$ energy flux per decade (same for $\nu_e$ and ${\bar
\nu}_{\mu}$) from GRB models with pre-ejected SNR shells. We assumed
$\sim 10^{-1}$ of protons escape the fireball shocks to reach the
shell, where $f_{pp} = 1$ and $E_{\nu}^{\rm s} = 10^9$ GeV.  The top
curve is calculated assuming all the GRBs have a SNR shell.  This is
an upper limit for maximum $\nu$'s through $pp$ and $p\gamma$
interactions in the SNR shell.  The bottom curve assumes 10\% of the
GRBs have an SNR shell.  Also shown are the diffuse $\nu$-fluxes from
$p\gamma$ interactions in the internal shocks of bursts (WB
\cite{wb97}, short dashed curve), and in GRB afterglows (\cite{wb00},
short dashed curve), as well as the $\nu$-flux obtained from the
cosmic ray limits (WB Limit \cite{wb99}, long dashed straight lines).
In a km$^2$ detector \cite{exp}, the number of diffuse events assuming
10\% of GRBs have SNR shells is $\sim$ 6 yr$^{-1}$ sr$^{-1}$ at
TeV-PeV, and $\sim$ 0.3 yr$^{-1}$ sr$^{-1}$ in the PeV-EeV range.
 
{\it Discussion and Implications.---} The single-burst $\nu$-fluxes
calculated here are predicated on the existence of a pre-ejected SNR
shell by the progenitor of the GRB, which occurs at the same location
after a delay of weeks \cite{vie98}. The $\nu$'s produced by $pp$ and
$p\gamma$ interactions between GRB relativistic protons and SNR shell
target protons and photons will be contemporary and of similar
duration as the GRB electromagnetic event. The high $pp$ optical depth
of the shell also implies a moderately high average Thomson optical
depth $\tau_{\rm T} \propto t_d^{-2}$ of the shell, dropping below
unity after $\sim 100$ d. Large scale anisotropy as well as clumpiness
of the shell will result in a mixture of higher and lower optical
depth regions being observable simultaneously, as required in the
supranova interpretation of X-ray lines and photon continua in some
GRB afterglows \cite{piro00}.  Depending on the fraction of GRB with
SNR shells, the contributions of these to the GRB diffuse $\nu$-flux
has a $pp$ component which is relatively stronger at TeV-PeV energies
than the internal shock $p\gamma$ component of \cite{wb97}, and a
shell $p\gamma$ component which is a factor 1 (0.1) of the internal
shock $p\gamma$ component (Fig. \ref{fig:difflux}) for a fraction 1
(0.1) of GRB with SNR shells. Due to a higher synchrotron cooling
break in the shell, at $E_\nu\gtrsim 10^{17}$ eV the shell component
could compete with the internal \cite{wb97} and afterglow \cite{wb00}
components.

Our $pp$ component is caused by internal shock-accelerated power-law
protons contemporaneous with the GRB event, differing from \cite{gg02}
who considered quasi-monoenergetic $\gamma_p\sim 10^{4.5}$ protons
from an MHD wind over $4\pi$ leading to a $\sim 10$ TeV neutrino
months-long precursor of the GRB. Our $p\gamma$ component arises from
the same GRB-contemporaneous internal shock protons interacting with
thermal 0.1 keV photons within the shell wall, whereas \cite{gg02}
consider such protons interacting with photons from the MHD wind
inside the shell cavity.

The pre-ejected supernova (supranova) model of GRB is a subject of
interest and debate \cite{m02} for interpreting the $\gamma$- and 
X-ray data, and independent tests would be useful. The
neutrino fluxes discussed here provide such a test, the predicted
event rates being detectable with kilometer scale planned Cherenkov
detectors.

SR thanks J.~P. Ralston for helpful discussions.  Work supported by
NSF AST0098416.

\end{document}